\documentclass{PoS}

\usepackage{amsmath}
\usepackage[percent]{overpic}

\title{Observations of the Pulsar Wind Nebula HESS\,J1825--137 with H.E.S.S. II}

\ShortTitle{Observations of the PWN HESS\,J1825--137 with H.E.S.S. II}

\author{\speaker{A.~M.~W. Mitchell}$^1$, S. Caroff$^2$, R.~D. Parsons$^1$, J. 
Hahn$^1$, V. Marandon$^1$ and J. Hinton$^1$ for the H.E.S.S. Collaboration\\

$^1$ Max-Planck-Institute f\"{u}r Kernphysik, Heidelberg, Germany\\
$^2$ Laboratoire Leprince-Ringuet, \'{E}cole Polytechnique, CNRS/IN2P3, Paris, France\\

 E-mail: \email{Alison.Mitchell@mpi-hd.mpg.de},\\
\email{sami.caroff@llr.in2p3.fr},\\
\email{Daniel.Parsons@mpi-hd.mpg.de},\\
\email{Joachim.Hahn@mpi-hd.mpg.de},\\
 \email{Vincent.Marandon@mpi-hd.mpg.de},\\
\email{Jim.Hinton@mpi-hd.mpg.de}
}

\abstract{We present a new and deep analysis of the pulsar wind nebula (PWN) HESS\,J1825--137 with a comprehensive data set of almost 400 hours taken with the H.E.S.S. array between 2004 and 2016. The large amount of data, and the inclusion of low-threshold H.E.S.S. II data allows us to include a wide energy range of more than 2.5 orders of magnitude, ranging from 150 GeV up to 70 TeV. We exploit this rich data set to study the morphology and the spectral distributions of various subregions of this largely extended source in more detail. We find that HESS\,J1825--137 is not only the brightest source in that region above 32 TeV, but is also one of the most luminous of all firmly identified pulsar wind nebulae in the Milky Way.}

\FullConference{35th International Cosmic Ray Conference --- ICRC2017\\
		10--20 July, 2017\\
		Bexco, Busan, Korea}

\begin{document}

\section{Introduction}
\label{sec:intro}

The pulsar wind nebula (PWN) HESS\,J1825--137 has been known as a strong TeV gamma-ray emitter since the first H.E.S.S. galactic plane survey in 2005 \cite{Aharonian05pwn1825}. It has become the archetypal example of energy dependent morphology in PWNe, due to the strength of this dependency seen in the nebula \cite{Funk06}. Towards high energies, the emission is seen to become more compact around the pulsar PSR\,B1823--13 (a pulsar with similar properties to the Vela pulsar) leading to an indisputable association of the nebula with this pulsar, and putting it at a distance of around 4 kpc \cite{atnf}. Conversely, at low energies HESS\,J1825--137 has one of the largest intrinsic physical extents of any known PWN. 

As has been previously presented \cite{gamma1825}, the treatment of the analysis of this nebula splits the data into two datasets, A and B, depending on whether or not the fifth and largest telescope of the H.E.S.S. array is included in the analysis. In both cases, stereoscopic events were required (i.e. more than one telescope triggering on any given air shower event). A sensitive likelihood-based analysis procedure, ImPACT (with a minimum image amplitude cut of 60 p.e.), was used for this purpose \cite{Parsons14}. The results presented in these proceedings have been cross-checked using an independent calibration and analysis chain based on an air-shower modelling approach \cite{deNaurois09}.

In this analysis, data taken during an observation campaign with HESS II in 2016 is included for the first time, increasing the livetime of the dataset with HESS II (dataset B) to 101 hours total, and the livetime for an analysis using the HESS I telescopes (dataset A) to 387 hours. Note that data contributing to dataset B (using telescopes CT1-5) may also contribute to dataset A (using telescopes CT1-4). 

\section{Properties of the PWN HESS\,J1825--137}
\label{sec_pwn}

\begin{figure}
\centering
\begin{overpic}[width=0.48\textwidth]{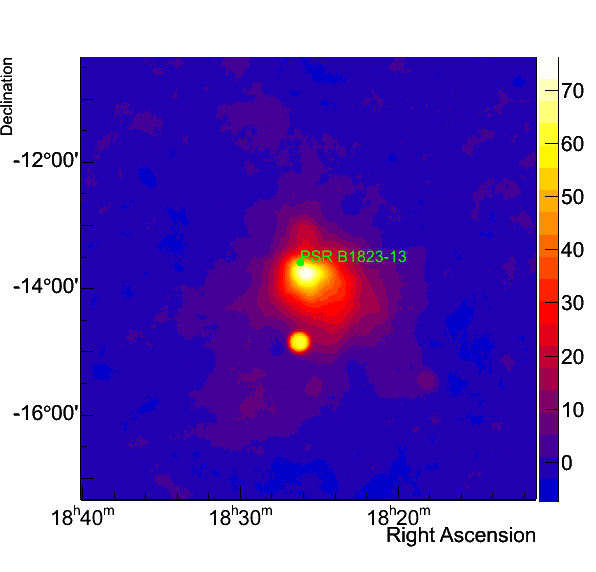}
\put(30,75){{\large{\textcolor{white}{A}}}}
\put(60,75){{\large{\textcolor{white}{\textit{preliminary}}}}}
\end{overpic}
\begin{overpic}[width=0.48\textwidth]{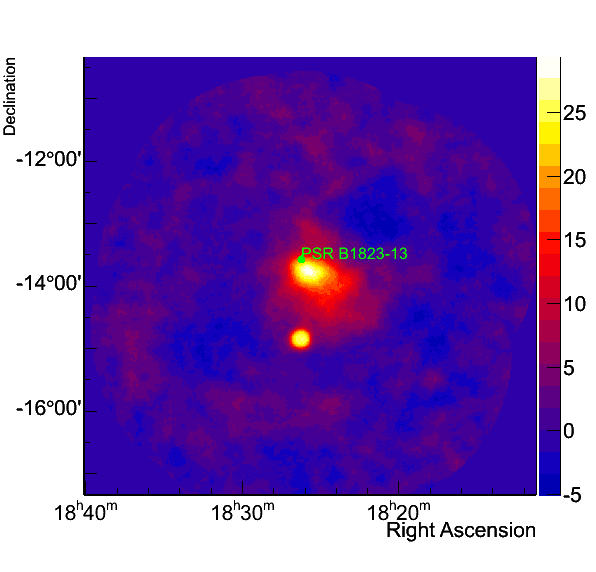}
\put(30,75){{\large{\textcolor{white}{B}}}}
\put(60,75){{\large{\textcolor{white}{\textit{preliminary}}}}}
\end{overpic}
\caption{Significance maps of HESS\,J1825--137 produced with HESS I (dataset A, left) and HESS II (dataset B, right) respectively.}
\label{fig:sigmaps}
\end{figure}

Significance maps of the sky region produced with both datasets are shown in Fig. \ref{fig:sigmaps}. Unsurprisingly, dataset A with the larger livetime has a much higher peak significance, and both datasets are dominated by the peak of the nebula emission close to (but offset from) the pulsar. Also visible within the field of view are two other known gamma-ray sources; the point-like binary system LS~5039 towards the south of the nebula, and the extended source HESS\,J1826--130, visible as a region of light blue towards the North of the pulsar \cite{ls5039,gamma1826}. This latter source becomes more prominent at higher energies, as the level of cross-contamination between the two sources decreases with increasing energy threshold. 

As can be seen in Fig. \ref{fig:sigmaps}, the nebula appears less peaked and more extended with dataset B. This is most likely due to the lower exposure time on this region, resulting in a lower sensitivity to the highest energy events from this region, which occur nearer to the pulsar. Additionally, a higher level of background systematics are visible, which is also most likely due to the lower overall exposure time with dataset B, as well as the smaller field of view of CT5, which renders the background estimation more complex. 

\begin{figure}
\centering
\begin{overpic}[width=0.48\textwidth]{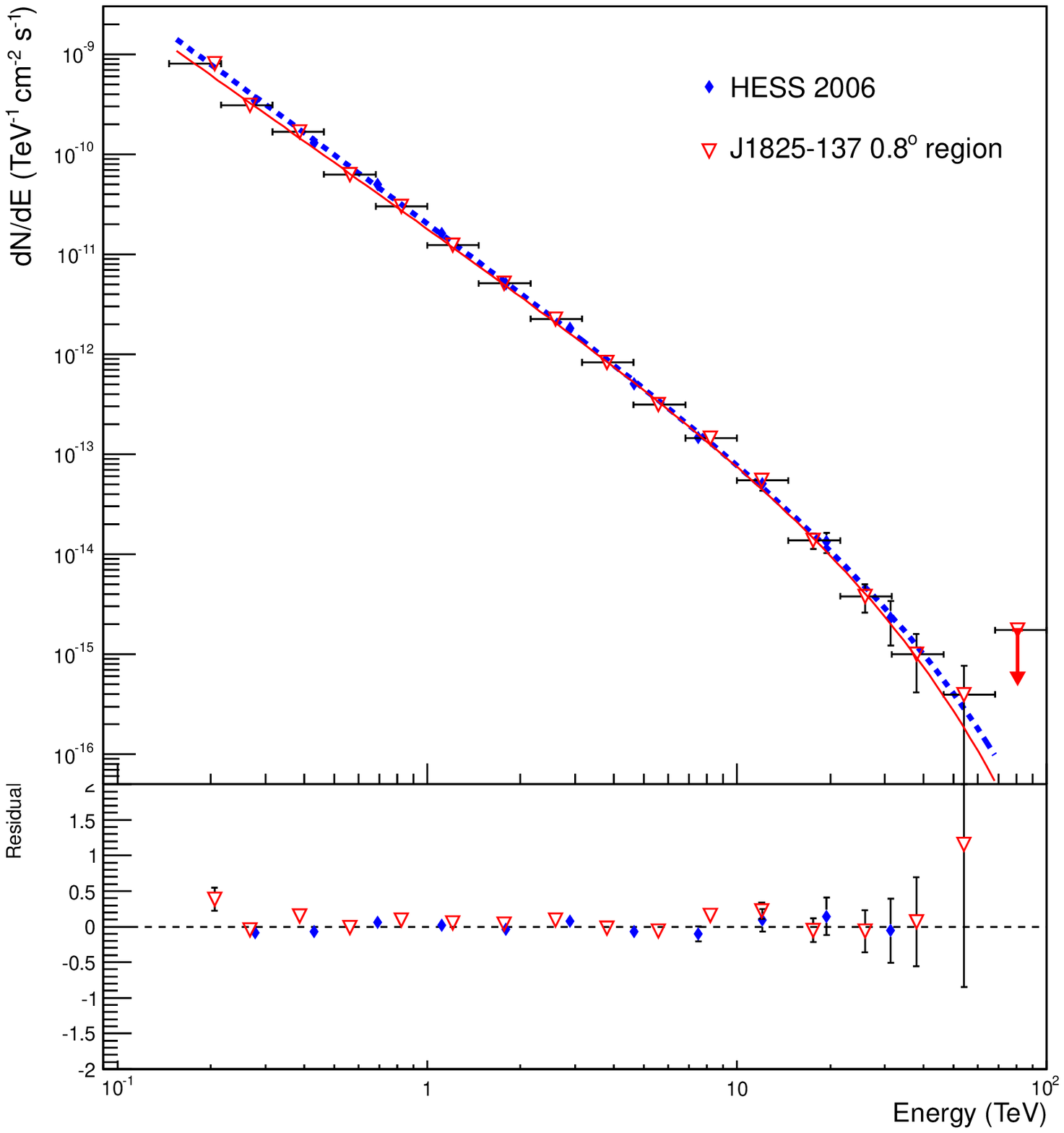}
\put(20,40){{\large{\textit{preliminary}}}}
\end{overpic}
\begin{overpic}[width=0.48\textwidth]{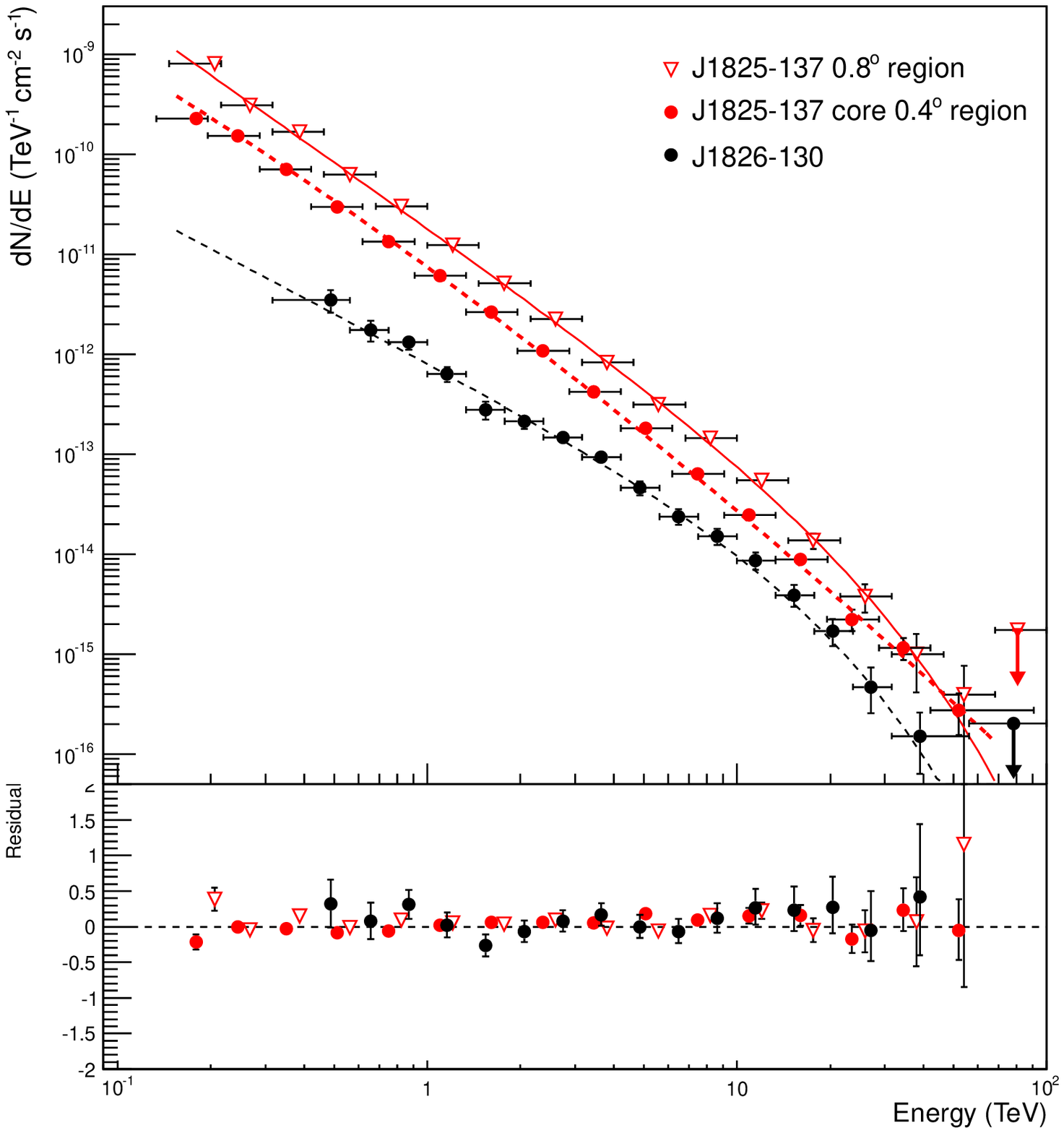}
\put(20,40){{\large{\textit{preliminary}}}}
\end{overpic}
\caption{Left: Differential Energy Spectrum extracted from the same region as was used in the 2006 HESS paper and shown in Fig. \ref{fig:highemap}. A power law with exponential cut-off fit model was used to calculate the residuals. Right: Differential Energy Spectrum extracted from $0.8^{\circ}$ region, compared to the core $0.4^{\circ}$ region and to HESS\,J1826--130. A log parabola model was favoured for the core region.}
\label{fig:specA06}
\end{figure}

\begin{table}
\centering
\begin{tabular}{lccc}
\hline 
$\mathrm{d}N/\mathrm{d}E$ Fit Model & $I_0$ (1 TeV) / $10^{-11}\mathrm{cm}^{-2}\mathrm{s}^{-1}$& $\Gamma$ & Fit Parameters \\
\hline
($0.8^{\circ}$): $I_0 E^{-\Gamma}$  & $1.79 \pm 0.02$ & $2.33 \pm 0.01$ & \\
($0.8^{\circ}$): $I_0 E^{-\Gamma}\exp (-E/E_c)$  & $1.88 \pm 0.02$ & $2.17 \pm 0.02$ & $E_c = 19 \pm 3 $ TeV \\
($0.8^{\circ}$): $I_0 E^{-\Gamma + \beta\log E}$  & $1.93 \pm 0.03$ & $2.31 \pm 0.01$ & $\beta = 0.076 \pm 0.009$ \\
\hline
(2006): $I_0 E^{-\Gamma}$ & $1.98 \pm 0.04$ & $2.38 \pm 0.02$ & \\
(2006): $I_0 E^{-\Gamma}\exp (-E/E_c)$ & $2.10 \pm 0.05$ & $2.26 \pm 0.03$ & $E_c = 24.8 \pm 7.2 $ TeV \\
(2006): $I_0 E^{-\Gamma + \beta\log E}$ & $2.10 \pm 0.04$ & $2.29 \pm 0.02$ &  $\beta = -0.17 \pm 0.04$\\
\hline
($0.4^{\circ}$): $I_0 E^{-\Gamma}$  & $0.681 \pm 0.007$ & $2.28 \pm 0.01$ & \\
($0.4^{\circ}$): $I_0 E^{-\Gamma}\exp (-E/E_c)$  & $0.720 \pm 0.009$ & $2.13 \pm 0.02$ & $E_c = 19 \pm 2 $ TeV \\
($0.4^{\circ}$): $I_0 E^{-\Gamma + \beta\log E}$ & $0.74 \pm 0.01$ & $2.26 \pm 0.01$ & $\beta = 0.078 \pm 0.008$ \\
\end{tabular}
\caption{Spectral fit parameters  for power law, power law with exponential cut-off and log parabola fit models, corresponding to a spectrum using dataset A, extracted from the $0.8^\circ$ radius region shown in Fig. \ref{fig:highemap}, as was used in the 2006 HESS\,J1825-137 analysis, as well as from the smaller $0.4^\circ$ radius region using the same central position.}
\end{table}

\begin{figure}
\centering
\begin{overpic}[width=0.45\textwidth]{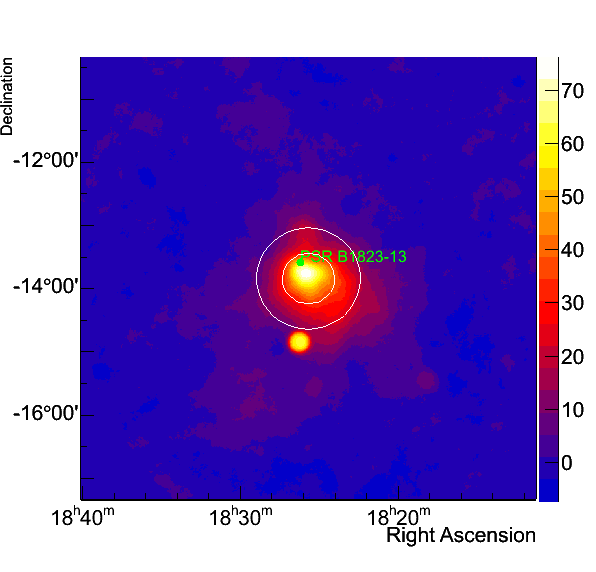}
\put(60,75){{\large{\textcolor{white}{\textit{preliminary}}}}}
\end{overpic}
\begin{overpic}[width=0.45\textwidth]{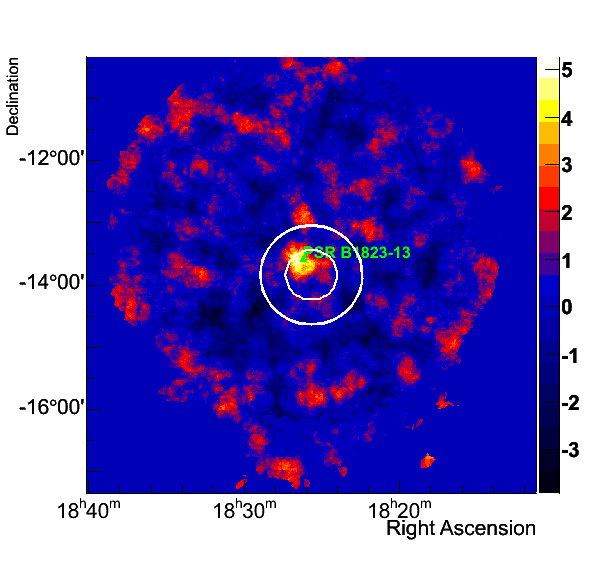}
\put(60,75){{\large{\textcolor{white}{\textit{preliminary}}}}}
\end{overpic}
\caption{Left: Significance map of the nebula over the full energy range showing the regions used for the spectra shown in Fig. \ref{fig:specA06}. Right: Significance map of the nebula at energies above 32~TeV. Both maps are made using dataset A.}
\label{fig:highemap}
\end{figure}

A spectrum was extracted from a region with $0.8^{\circ}$ radius as defined in the 2006 HESS\,J1825-137 analysis, based on the significance contours of the region at that time, centred on the mean of a 2D Gaussian morphology fit \cite{Funk06}. The comparison of this spectrum with that of the 2006 HESS\,J1825-137 analysis is shown in Fig. \ref{fig:specA06}. In the intervening years, the presence of HESS\,J1826--130 as an independent source has been confirmed, with which this spectral extraction region overlaps \cite{gamma1826}. 
In order to be certain that a spectrum is extracted from HESS\,J1825--137 alone, a complementary spectrum was extracted from a core region of the nebula with $0.4^{\circ}$ radius, using the same central position as for the $0.8^{\circ}$ radius region. These two spectral extraction regions are shown overlaid on a significance map of the region in Fig. \ref{fig:highemap}. Fig. \ref{fig:specA06} compares the spectra extracted from these two regions with the spectrum for HESS\,J1826--130 \cite{icrc1826}. It can be seen that towards the highest energies, the core region can account for the entire flux of the $0.8^{\circ}$ spectral region, dominating over any contribution from the overlap with HESS\,J1826--130. 
This spectral analysis also shows that HESS\,J1825--137 is also one of the most luminous of all firmly identified PWNe in the Milky Way \cite{Klepser17}.

In order to help identify the origin of the highest energy emission, it is helpful to study the region at higher energies where the two sources become more significantly separated. Fig. \ref{fig:highemap} also shows a significance map of the region with a minimum energy cut of 32~TeV. At these energies, it can be seen that the peak of the emission from HESS\,J1825--137 appears shifted towards the pulsar PSR~B1823-13 (with respect to the peak emission over all energies), and that emission from HESS\,J1825--137 is more significant than that from HESS\,J1826--130.

\begin{figure}
\centering
\begin{overpic}[trim= 0mm 0mm 0mm 2cm,clip,width=0.45\textwidth]{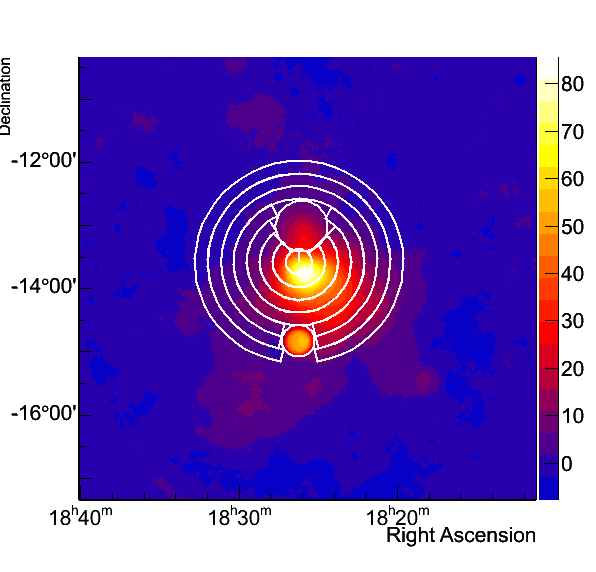}
\put(60,75){{\large{\textcolor{white}{\textit{preliminary}}}}}
\end{overpic}
\caption{Excess map of the nebula showing rings and masked regions used for extracting the azimuthal profiles, out to a radial distance of 1.6$^{\circ}$ from the pulsar.}
\label{fig:specoverlay}
\end{figure}

\section{Determining the Major Axis}
\label{sec:major}

\begin{figure}
\centering
\begin{overpic}[trim= 0mm 0mm 0mm 1.5cm,clip,width=0.45\textwidth]{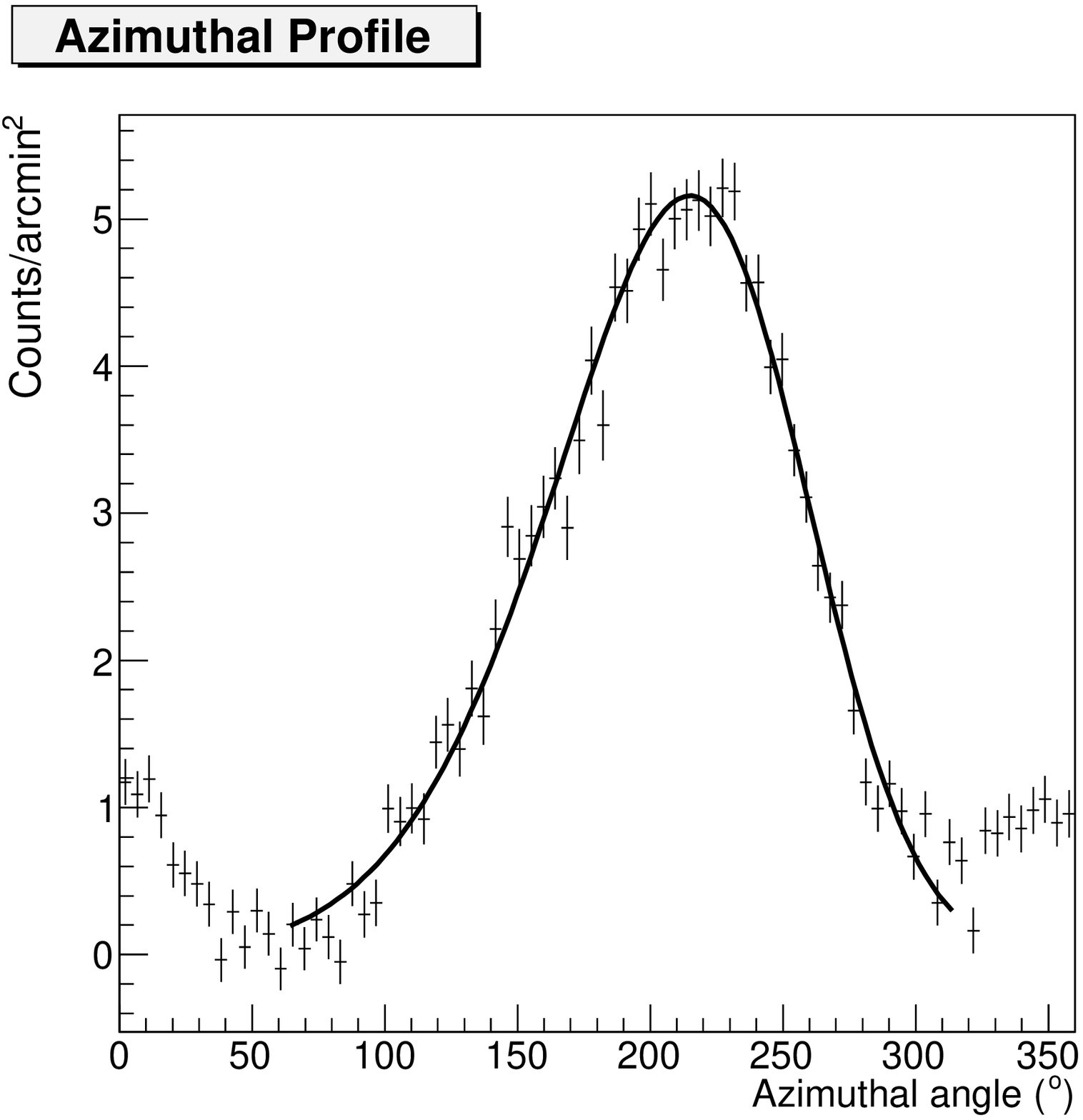}
\put(20,75){{\large{\textit{preliminary}}}}
\put(80,75){{\large{A}}}
\end{overpic}
\begin{overpic}[trim= 0mm 0mm 0mm 1.5cm,clip,width=0.45\textwidth]{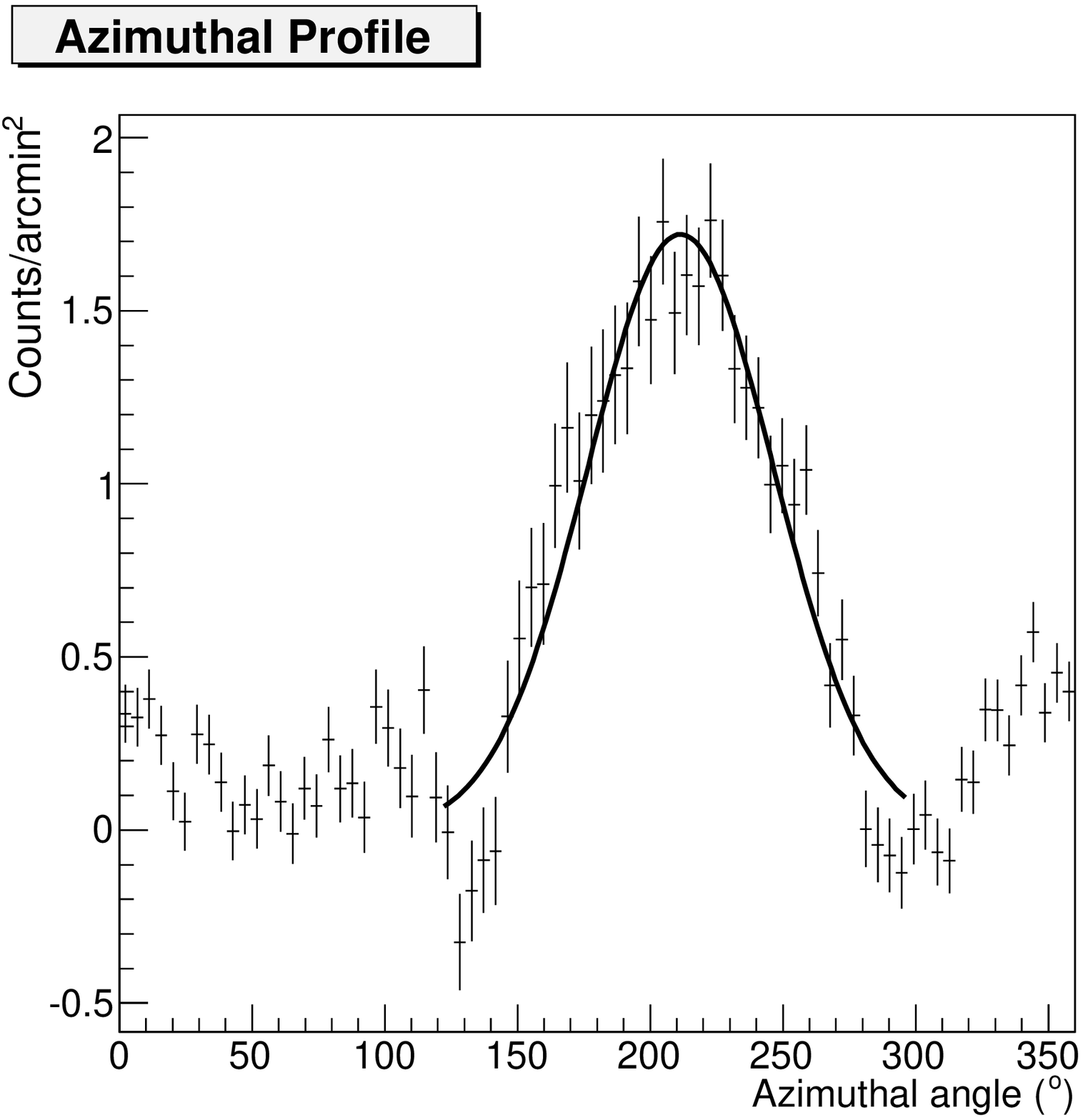}
\put(20,75){{\large{\textit{preliminary}}}}
\put(80,75){{\large{B}}}
\end{overpic}
\caption{Azimuthal profiles extracted from the excess map shown in Fig. \ref{fig:specoverlay}, from all regions out to a radial distance of 1.6$^{\circ}$ from the pulsar. Left: dataset A, Right: dataset B. }
\label{fig:aziprofile}
\end{figure}

In order to determine the primary direction of the nebula emission, the following procedure was adopted. From the excess map of the HESS\,J1825--137 region, a series of concentric ring shaped regions were defined, of $0.2^{\circ}$ width, centered on the pulsar PSR\,B1823--13. The excess counts were extracted from each of these regions separately, to produce a series of one dimensional profiles of the emission as a function of azimuthal angle around the pulsar. The direction of positive declination was arbitrarily chosen to correspond to $0^{\circ}$ in azimuth angle, increasing in the anticlockwise direction as seen on the excess map in Fig. \ref{fig:specoverlay}. 
Masks were applied, over the two other sources in the region; HESS\,J1826--130 (with a mask of $0.4^{\circ}$ radius) and LS\,5039 (with a mask of $0.25^{\circ}$ radius) such that emission from these regions was not included in the excess profile.
A moving average procedure was then used to find the peak of the emission; and similarly to find two minima, within $180^{\circ}$ regions either side of the peak. The excess profile was then fit with a Gaussian function between these minima in each radial band. The mean of the Gaussian fits to all excess profiles was found to shift gradually with increasing radial distance. The same procedure was used to fit the excess profile extracted from all regions out to $1.6^{\circ}$ distance from the pulsar, as shown in Fig. \ref{fig:aziprofile}.


The mean of the Gaussian fit to an azimuthal profile of the full region, as shown in Fig. \ref{fig:aziprofile}, was found to be compatible with the average of the means of the Gaussian fits to the separated radial bands. Consequently, the major axis of the emission was found to be orientated along $214^\circ \pm 4^\circ \mathrm{stat.} \pm 4^\circ \mathrm{sys.}$, where the statistical errors correspond to those obtained from the fit. The systematic error is taken from the differences between the two analysis pipelines. 

From this definition of the major axis, the minor axis was simply defined as being orthogonal to this axis. 

\section{Determining the Nebula Extent}
\label{sec:extent}

The minor axis was then used to define a semicircular region, centered on the pulsar, within which the radial extent of the nebula could be characterised. In a similar manner to the azimuthal profiles, the excess counts were extracted from this region, and a radial profile constructed, clearly showing decreasing nebula emission with increasing distance from the pulsar. However, the peak of the nebula emission can be seen to be offset from the pulsar position towards the South. Again, the binary gamma-ray source LS\,5039 was masked in the extraction of counts from the excess map, such that it does not contribute to the radial profile. 

\begin{figure}q
\centering
\begin{overpic}[width=0.75\textwidth]{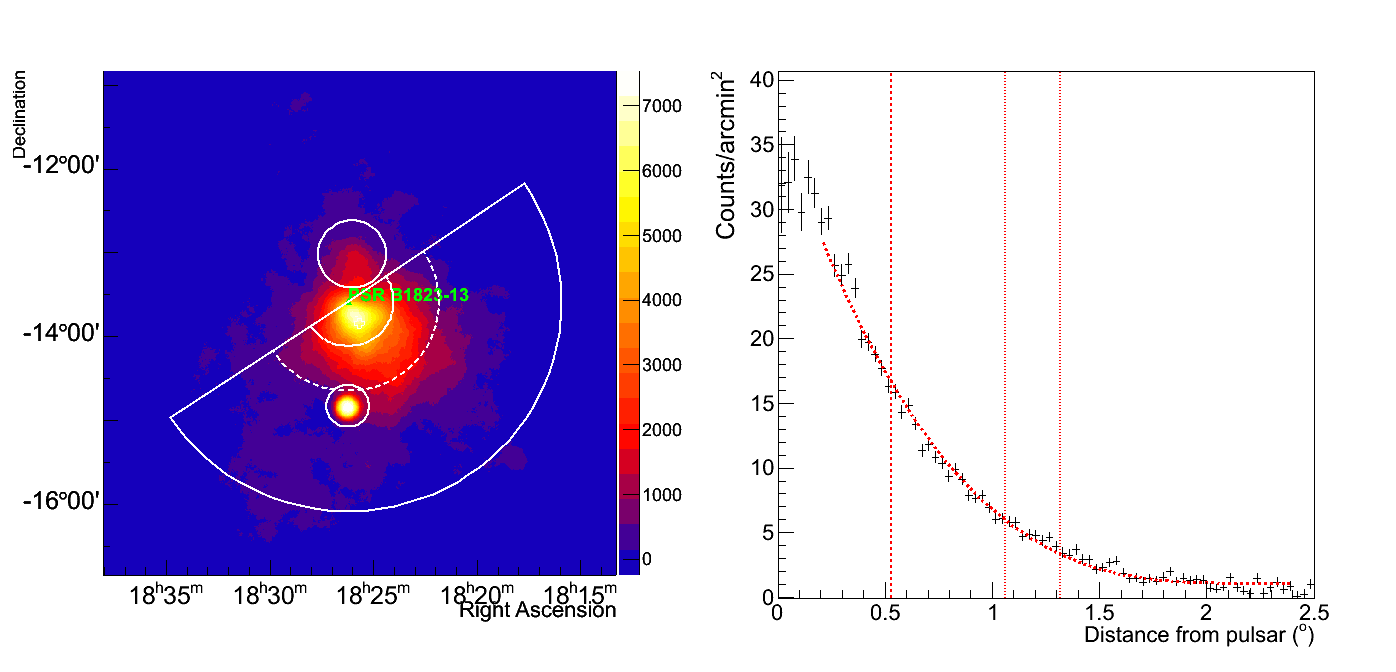}
\put(10,38){{\large{\textcolor{white}{\textit{preliminary}}}}}
\end{overpic}
\begin{overpic}[width=0.75\textwidth]{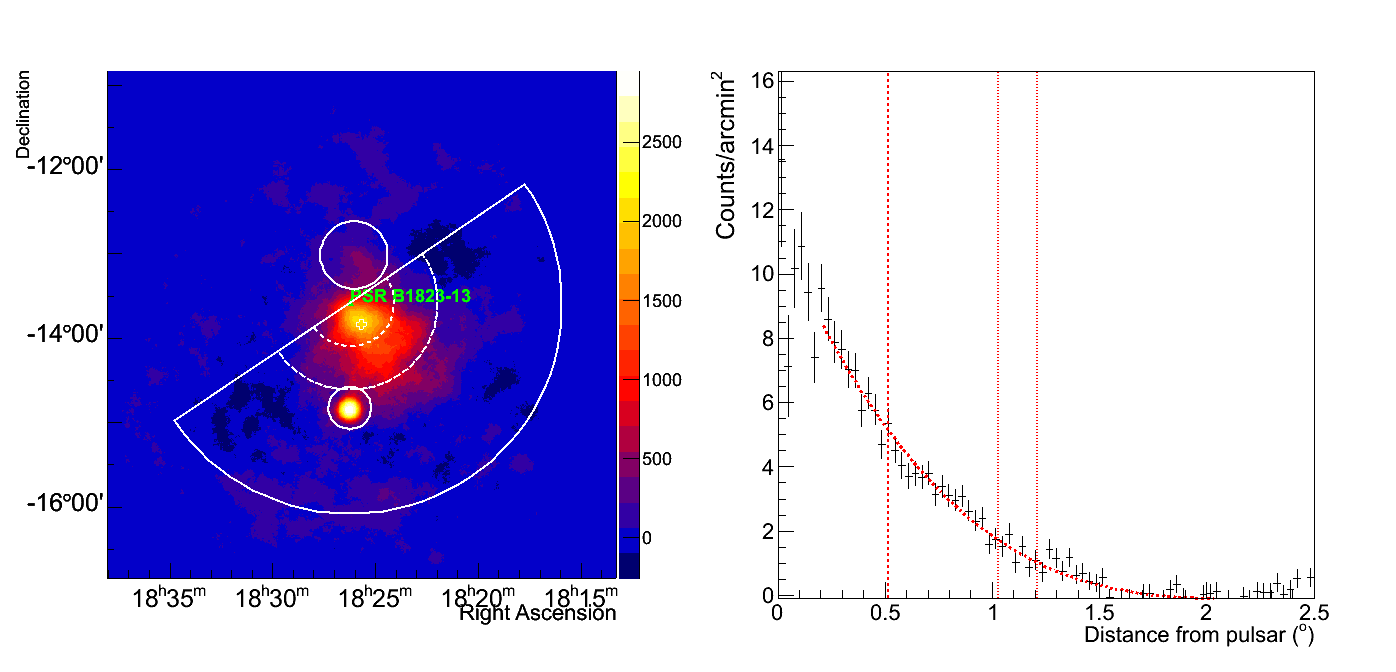}
\put(10,38){{\large{\textcolor{white}{\textit{preliminary}}}}}
\end{overpic}
\caption{Radial profiles and extraction region used for measuring the extent. Above - dataset A; Below - dataset B. The upper and lower circles correspond to the exclusion regions around the sources HESS\,J1826--130 and LS\,5039.}
\label{fig:radprofs}
\end{figure}

In order to characterise this drop off of the emission with increasing distance, the radial profile (as shown in Fig. \ref{fig:radprofs}) was fit with the following polynomial function:

\begin{equation}
y = 
\begin{cases}
a(r-r_0)^n +c & ( x < r_0 ) \\
c & ( x \geq r_0 )
\end{cases}~,
\label{eq:polfit}
\end{equation}
such that with increasing $r$, the emission decreases out to a distance $r_0$ at which it assumes a constant value, $c$. 
The parameter $a$ provides the overall normalisation, whilst the parameter $r_0$ was found to be highly sensitive to the value of $n$, the order of the polynomial.  

Instead of using the parameter $r_0$ as a direct measure of the extent, a moving average was used to find the location and value of the peak of the emission. The radius at which the fitted function dropped to a fixed fraction of the peak value (e.g. $50\%$, $r_{50}$), was then used as an alternative measure of the nebula extent. In contrast to $r_0$, the parameter $r_{50}$ was found to be robust against the order of the polynomial $n$. This measured radius, $r_{50}$, may be considered as roughly equivalent to a Half Width at Half Maximum measure. 

Fig. \ref{fig:radprofs} shows both the region extracted from the excess map and the fitted radial profile for datasets A and B. The minimum radius of the fit was $0.2^{\circ}$; this choice was found not to affect the fit results provided the minimum radius is beyond the peak. Vertical dashed lines on the radial profiles illustrate the approximate location of $r_{50}$, twice $r_{50}$ and $r_{10}$, (the radius at which the flux drops to $10\%$ of the peak) respectively. 
The extents obtained in this manner were cross-checked through fitting an exponential function to the radial profile instead of the polynomial (equation \ref{eq:polfit}), and the evaluated $r_{50}$ obtained from the fitted function were found to be consistent. Performing these fits multiple times for different major axes, as determined by the two analysis pipelines, enabled a systematic error to be derived on the value of $r_{50}$ obtained. 

A toy Monte Carlo technique was used to determine the statistical errors on the fitted extents. For dataset A, a value of $r_{50} = 0.53^\circ \pm 0.02^\circ \mathrm{stat.} \pm 0.06^\circ \mathrm{sys.}$ was found; whilst for dataset B $r_{50} = 0.51^\circ \pm 0.04^\circ \mathrm{stat.} \pm 0.06^\circ \mathrm{sys.}$, well within the errors of the technique. 


\section{Conclusion}
\label{sec:conclude}

In this proceedings, we have endeavoured to extract further properties of the PWN HESS\,J1825--137, characterising the nature of the emission and overall nebula extent in more depth with this rich dataset than has previously been shown. 
Of particular note should be the energies which are reached in the innermost part of the nebula. With significant data up to several tens of TeV and an exponential cut-off of around 20~TeV, HESS\,J1825--137 is at least as powerful a TeV accelerator as its neighbour HESS\,J1826--130, previously suggested as a PeVatron candidate, yet with an exponential cut-off of $\sim$13~TeV. By extracting spectra from spatially resolved regions across the nebula, it has been demonstrated that the high energies reached in the full nebula spectrum of Fig. \ref{fig:specA06} are not due to contamination by HESS\,J1826--130, but are attributable to HESS\,J1825--137 alone. This spectral analysis also shows that HESS\,J1825--137 is also one of the most luminous of all firmly identified PWNe in the Milky Way. 

A precise morphological study has been conducted, made possible thanks to the rich VHE data set available on the source. Such an approach towards characterising the extent may be used in order to probe the dependence of the nebula extent on energy, by measuring how the fitted value of $r_{50}$ varies with energy band of the excess distribution. 
This, and further studies into the nebula properties, are left to a future H.E.S.S. publication with this rich dataset.

\section*{Acknowledgements}

\textit{\small{The support of the Namibian authorities and of the University of Namibia in facilitating the construction and operation of H.E.S.S. is gratefully acknowledged, as is the support by the German Ministry for Education and Research (BMBF), the Max Planck Society, the German Research Foundation (DFG), the Alexander von Humboldt Foundation, the Deutsche Forschungsgemeinschaft, the French Ministry for Research, the CNRS-IN2P3 and the Astroparticle Interdisciplinary Programme of the CNRS, the U.K. Science and Technology Facilities Council (STFC), the IPNP of the Charles University, the Czech Science Foundation, the Polish National Science Centre, the South African Department of Science and Technology and National Research Foundation, the University of Namibia, the National Commission on Research, Science \& Technology of Namibia (NCRST), the Innsbruck University, the Austrian Science Fund (FWF), and the Austrian Federal Ministry for Science, Research and Economy, the University of Adelaide and the Australian Research Council, the Japan Society for the Promotion of Science and by the University of Amsterdam.
We appreciate the excellent work of the technical support staff in Berlin, Durham, Hamburg, Heidelberg, Palaiseau, Paris, Saclay, and in Namibia in the construction and operation of the equipment. This work benefited from services provided by the H.E.S.S. Virtual Organisation, supported by the national resource providers of the EGI Federation.}}

\end{document}